\documentclass[11pt]{article}
\usepackage{amsfonts}
\usepackage{amsfonts}
\usepackage{amsfonts,amsmath,amssymb,epsf}
\usepackage{graphicx,color}
\usepackage[usenames,dvipsnames]{pstricks}
\usepackage{epsfig}
\usepackage{pst-grad} 
\usepackage{pst-plot} 
\usepackage{verbatim}
\newcommand{\be}{\begin{equation}}
\newcommand{\ba}{\begin{eqnarray}}
\newcommand{\ea}{\end{eqnarray}}
\newcommand{\ee}{\end{equation}}

\newcommand{\f}{\frac}

\begin{document}

\begin{titlepage}
\thispagestyle{empty}

\begin{flushright}
KUNS-2188
\end{flushright}

\bigskip

\begin{center}
\noindent{\Large \textbf{BPS operators from the Wilson loop in the 3-dimensional supersymmetric Chern-Simons theory}}\\
\vspace{15mm} Mitsutoshi Fujita\footnote{email: mfujita@gauge.scphys.kyoto-u.ac.jp} \\
\vspace{1cm}

 {\it Department of Physics, Kyoto University, Kyoto 606-8502, Japan \\}

\vskip 3em
\end{center}

\begin{abstract}
We consider the small deformation of the point-like Wilson loop in the 3-dimensional $\mathcal{N}=6$ superconformal Chern-Simons theory. By Taylor expansion of the point-like Wilson loop in powers of the loop variables, we obtain the BPS operators that correspond to the excited string states of the dual IIA string theory on the pp wave background. The BPS conditions of the Wilson loop constrain both the loop variables and the forms of the operators obtained in the Taylor expansion.
\end{abstract}
\end{titlepage}

\newpage

\section{Introduction}
The dual gravity interpretation of the supersymmetric Wilson loop in the $D=4$ $\mathcal{N}=4$ SYM theory has  been important in the context of the AdS/CFT correspondence \cite{BA1}. The supersymmetric Wilson loop contains the 6 scalar fields $\Phi^i$ and the expectation value of the Wilson loop is protected from the UV divergence \cite{Dru02}.
 The expectation value of the circular Wilson loop was obtained by analyzing the string minimal surface in the anti de Sitter space \cite{BS1}. Furthermore,  the supersymmetric Wilson loop in the BMN sector can be described by the dual IIB string theory on the pp wave background. The BMN correspondence \cite{BP1} is the duality between 
 the infinite strings of operators in $\mathcal{N}=4$ SYM and the excited string states in the dual IIB string theory on the pp wave background.
 In \cite{BD2}, it was conjectured that the 1/2-BPS point-like Wilson loop $W(C_0)$ for $C_0$ shrinking to a spacetime point is mapped to the vacuum state of the dual IIB string theory on the pp wave background. The functional derivatives of the Wilson loop are mapped to the excited string states. We have summarized the results obtained in \cite{BD2} in Table 1.
\vskip2mm
 \begin{center}
\begin{tabular}{cc|c} \hline
 \multicolumn{2}{c|}{SYM side}  & dual IIB string side  \\ \hline
 functional derivatives & corresponding operators &   \\ \hline
$-$ & $W(C_0)$ & $|0;p^+>$ \\
 $\int dse^{\f{2\pi ins}{t}}\f{\delta}{\delta x^{\mu}(s)}$ & $D_{\mu}Z$ & $\alpha ^{\mu}_{(n)}$\\
$\int dse^{\f{2\pi ins}{t}}\f{\delta}{\delta y^a(s)}$ & $\Phi^a$ & $\alpha ^{4+a}_{(n)}$ \\ \hline
\end{tabular}
\vskip2mm 
Table 1: The relation between the functional derivatives and the IIB string oscillation modes $\alpha^{\mu}_{(n)}$ and $\alpha^{4+a}_{(n)}$. In the table, $\mu=0,1,2,3$, $a=1,2,3,4$ and $Z=\Phi^5+i\Phi^6$. 
\end{center}
\vskip2mm

In this paper, motivated by the work \cite{BD2}, we consider the dual IIA string theory description of the supersymmetric Wilson loop in the recently proposed $D=3$ $\mathcal{N}=6$ Chern-Simons-matter theory (ABJM theory) \cite{ABJM,ABJM2}. The ABJM theory  is the low-energy effective theory of the $N$ M2-branes at the singularity of the orbifold $\mathbb{C}^4/\mathbb{Z}_k$.\footnote{In the special case of the gauge group $SU(2)\times SU(2)$, the ABJM theory has the $SO(8)$ enhanced global $R$-symmetry.} This theory can be analyzed by using the dual IIA string theory on the $AdS_4\times CP^3$ spacetime and on its Penrose limit  \cite{BB1}-\cite{BB4} in the parameter regime 
\ba
\sqrt{\lambda}\gg 1,\quad e^{2\phi}\sim \dfrac{N^{\frac{1}{2}}}{k^{\frac{5}{2}}}\ll 1, \label{PAR11}
\ea 
where $k$ is the Chern-Simons level, $N$ is the rank of the gauge group, $\lambda =N/k$ is 't Hooft coupling and $\phi$ is the dilaton. The first one in \eqref{PAR11} implies that in the dual type IIA string theory, the radius of curvature is much larger than 1 in the string unit and the second one implies that we take the small string coupling limit, to suppress the quantum corrections. 

The supersymmetric Wilson loop in the ABJM theory was proposed in the literature \cite{BW1,Dru1,BW2}. The Wilson loop contains a product of the bi-fundamental scalars on the exponent.
It was shown that the straight line and circular Wilson loops preserve 1/6 of the ABJM supersymmetry. 

The main purpose of our paper is to study the dual IIA string theory description of the point-like Wilson loop that has enhanced 1/3-supersymmetry. 
 We show that by deforming the point-like Wilson loop, we can obtain the BPS operators that correspond to the excited string states of the dual IIA string theory on the pp wave background.\footnote{See \cite{BD1} on the deformation of the Wilson loop operator in the $\mathcal{N}=4$ SYM as well as in the YM thoeory.} The BPS conditions of the Wilson loop, \eqref{B313} and \eqref{B314}, give the constraint on both the loop variables and the forms of the BPS operators.
 By following the conjecture in \cite{BD2}, we give maps from the functional derivatives of the Wilson loop to the dual IIA string excited states.

 The Penrose limit of the dual gravity theory is given by the following limit:
 \begin{eqnarray}
 N,J\to \infty \ \text{with} \ \lambda^{\prime}=\dfrac{\lambda}{J^2} \  \text{fixed},
 \end{eqnarray}
 where $J$ is the charge of the infinite strings of operators under the $U(1)$ subgroup of $SU(4)$ R-symmetry. By determining the function $h(\lambda)$ that appear in the dispersion relation, the gauge/gravity correspondence has been proved up to the curvature corrections to the pp wave background \cite{BB0}.
 
The content of this paper is as follows: in section 2, we consider the point-like Wilson loop in the ABJM theory and compare it with the vacuum state of the dual IIA superstring theory. 
 In section 3, we obtain the BPS conditions for the Wilson loop. We show that the point-like Wilson loop is 1/3 BPS and the supersymmetry generator preserved by the point-like Wilson loop is the same as that preserved by the infinite chain dual to the IIA string vacuum state.
 In section 4, we solve the BPS equations satisfied by the loop variables and expand the Wilson loop in powers of the independent loop variables. Thus, we obtain the maps from the functional derivatives of the Wilson loop to the dual IIA string excited states.
 
\section{The supersymmetric Wilson loop in the ABJM theory}

The ABJM theory is the 3-dimensional $\mathcal{N}=6$ supersymmetric Chern-Simons theory with the gauge group $U(N)\times U(N)$. The fields in the ABJM theory are the
$U(N)\times U(N)$ gauge fields $A_{m}$ and $\hat{A}_{m}$, the bi-fundamental bosonic fields $Y^{I}$ ($Y^I=(A_1,A_2,\bar B_1,\bar B_2)$) and the bi-fundamental spinors $\psi _{I\alpha}$, where $I$ ($I=1,..,4$) is the index of the $SU_R(4)$ $R$-symmetry and $\alpha$ ($\alpha =1,2$) is the (2+1)-dimensional spinor index.
 
The Wilson loop in the ABJM theory \cite{BW1} was given by\footnote{In \cite{Dru1}, they also obtain the Wilson loop with the gauge field $\hat A_{m}$. We do not consider the Wilson loop with $\hat A_{m}$ here; our Wilson loop breaks the parity symmetry of ABJM theory (see \cite{BB3,Dba1}).} 
\ba W[C]=\mbox{Tr}\left[P\exp i\oint _Cds \left( \dot x^m(s) A_m+M_I{}^{J}(s) Y^IY_J^{\dagger}\right) \right], \label{W21} \ea
where $x^m(\tau)$ describes the path $C$ on $R^{1,2}$ and the function $M_I{}^J(s)$, determined by the SUSY, will be the coordinate of the transverse space $\mathbb{C}^4/\mathbb{Z}_k$.
We assume that $M_I{}^J(s)$ is a $4\times 4$ real matrix.\footnote{We can give $M_I{}^J(s)$ a $U_R(1)$ charge, which is a subgroup of the $SU_R(4)$ $R$-symmetry.
In the Wilson loop \eqref{W21}, the $U_R(1)$ symmetry that rotates $A_1$ and $B_1$ by $\alpha=\exp(i\varphi/2)$ 
 also operates on $M_I{}^J(s)$ as follows:
\ba &V^{-1}MV=\begin{pmatrix} \alpha &0&0&0 \\
0&1&0&0 \\
0 & 0 & \bar\alpha & 0 \\
0 & 0 & 0 & 1 \end{pmatrix} \begin{pmatrix} m_{11}&m_{12}&m_{13}&m_{14} \\
m_{21} &m_{22} &m_{23} & m_{24} \\
m_{31}&m_{32}&m_{33}& m_{34} \\
m_{41} &m_{42}&m_{43}&m_{44} \end{pmatrix}\begin{pmatrix} \bar\alpha &0&0&0 \\
0&1&0&0 \\
0 & 0 & \alpha & 0 \\
0 & 0 & 0 & 1 \end{pmatrix}\notag \\
&=\begin{pmatrix} m_{11}&\alpha m_{12}&\alpha ^2 m_{13}&\alpha m_{14} \\
\bar\alpha m_{21} &m_{22} &\alpha m_{23} & m_{24} \\
\bar\alpha ^2 m_{31}&\bar\alpha m_{32}&m_{33}&\bar\alpha m_{34} \\
\bar\alpha m_{41} &m_{42}&\alpha m_{43}&m_{44} \end{pmatrix}. 
 \notag \ea
In section 4, we use this $U_R(1)$ symmetry to classify the loop variables. }
 
We consider the point-like Wilson loop whose path $C_0$ shrinks to the point $x^m=x^m_0$ ($\dot{x}^m=0$). We set $M^I{}_J$ in a nilpotent matrix, 
\ba \begin{pmatrix} 0 & 0 & 1 & 0 \\
0 & 0& 0 & 0 \\
0 & 0& 0 & 0 \\
0 & 0& 0 & 0\end{pmatrix}. \label{M22} \ea
By expanding the exponential part of the Wilson loop, we obtain the infinite sum of the local operator as follows:
\ba W[C_0]=\mbox{Tr}[\exp\left( itA_1B_1(x_0)\right)]=\sum_{J=0}^{\infty}\dfrac{(it)^J}{J!}\mbox{Tr}[(A_1B_1)^J](x_0), \label{W23} \ea
where $t$ describes the periodicity of the loop; we identify $s=0$ with $s=t$. 
In the higher order of $J$, \eqref{M22} includes the infinite strings of the operator $A_1B_1$ that correspond to the vacuum state of the dual IIA string theory. 
 In section 4, by deforming \eqref{W23}, we also obtain the BPS operators that correspond to the dual IIA string excited states.
 
In the next section, we analyze the SUSY preserved by the Wilson loop \eqref{W21} and \eqref{W23}.

\section{Supersymmetry} 

The $\mathcal{N}=6$ SUSY generator described by the superspace coordinate was $\omega _{IJ}$
\footnote{$\omega _{IJ}$ is obtained by using the Clebsch-Gordan decomposition of the 6 Majorana spinors $\epsilon _i$ $(i=1,..,6)$, which are also the $\mathcal{N}=6$ SUSY generators (see Appendix B).}, which transforms as the anti-symmetric representation of $SU_R(4)$ and satisfies the following relations:

\ba &\left(\omega_{IJ,\alpha}\right)^*=\omega^{IJ}_{\alpha}, \quad \omega ^{IJ}_{\alpha}=\dfrac{1}{2}\epsilon ^{IJKL}\omega _{KL,\alpha}, \label{C32} \\
&\omega^{IK}\omega_{KJ}=\delta ^I{}_J\epsilon^i\epsilon_i, \label{C33} \ea
where $\epsilon^i$ $(i=1,...,6)$ are the Majorana spinors, which are also the $\mathcal{N}=6$
 SUSY generator.
The $\mathcal{N}=6$ SUSY transformations were given by
\ba
\delta Y^I &=& i \omega^{IJ} \psi_J,  \\
\delta Y_I^\dagger &=& i \psi^{\dagger \, J} \omega_{IJ},  \\\delta A_m &=& - (
Y^I \psi^{J \dagger} \gamma_m \omega_{IJ} 
+ \omega^{IJ} \gamma_m \psi_I Y_J^\dagger), \\
\delta \hat{A}_m &=& 
\psi^{I \dagger} Y^J \gamma_m \omega_{IJ} 
+ \omega^{IJ} \gamma_m Y_I^\dagger \psi_J, 
\label{S21} \ea
where the convention of the spinors is the same as given in \cite{ABJM2, BJ1}.

We consider the SUSY transformation of the Wilson loop as follows:
\ba e^{i(\omega _{IJ}Q^{IJ}+\omega ^{IJ}\bar Q {IJ})}W[C]e^{-i(\omega _{IJ}Q^{IJ}+\omega ^{IJ}\bar Q_{IJ})}=W[C]+\delta W[C], \ea
where $Q^{IJ}$ is the SUSY generator.
The Wilson loop preserves a part of the SUSY in the ABJM theory when $\delta W[C]=0$ for the arbitrary $s$. 

By using the condition $\delta W[C]=0$, we can show that the SUSY generator $\omega _{AB}$ preserved by the Wilson loop satisfies the following equations 
 as given in \cite{BW1}:
 \ba &\omega ^{\alpha}_{AB}\gamma _{m\alpha\beta}\dot x^m-iM_B{}^K\omega _{KA,\beta}=0, \label{E313} \\
&\omega ^{AB,\alpha}\gamma _{m\alpha\beta}\dot x^m-iM_I{}^A\omega ^{IB}_{\beta}=0.  
\label{E314} \ea
Note that the complex conjugate of \eqref{E313} gives \eqref{E314} when $M_I{}^J$ is a Hermite matrix.
When we contract these two equations by using $\epsilon _{\alpha\beta}$ and the charge conjugation matrix $\hat C$ (see appendix A) or when we multiply
 \eqref{E313} by $\gamma_n\dot x^n-iM^T$ from the right,\footnote{When we multiply \eqref{E314} by $\gamma_n\dot x^n-iM$ from the right, we obtain the same equation.} we obtain the following BPS conditions:
\ba &4\dot{x}^2(s)+M_I{}^K(s)M_{K}{}^I(s)=0,  \label{B313} \\
&\det(\dot{x}^2(s)+M^T(s)M^T(s))=\det(\dot{x}^2(s)+M(s)M(s))=0. \label{B314}
 \ea
\eqref{B313} and \eqref{B314} are the necessary condition to preserve a part of SUSY: a straight line and \eqref{W23} satisfy \eqref{B313}.  
\eqref{B313} is similar to the BPS conditions $\dot{x}^2+\dot{y}^2=0$ satisfied by the Wilson loop in the $d=4$, $\mathcal{N}=4$ SYM.

We can show easily that the point-like Wilson loop \eqref{W23} preserves 1/3 of the SUSY in the ABJM theory. For $\dot x_m=0$, the equations \eqref{E313} and \eqref{E314}
are given by
 \ba &M_B{}^K\omega _{KA,\beta}=0, \label{P316} \\
&M_I{}^B\omega ^{IA}_{\beta}=0.  \label{P317} \ea
By solving these equations, we obtain $\omega _{13}=\omega _{23}=\omega _{24}=\omega_{34}=0$.\footnote{The superconformal group satisfied by $\eqref{W23}$ is the $SU(2|2)$ generated by $\omega _{12}$ and $\omega _{14}$ as given in \cite{BB2}.} The point-like Wilson loop preserves the SUSY $\omega _{12}$ and $\omega _{14}$ which are not constrained by \eqref{P316} and \eqref{P317}. Note that the Wilson loop over the straight path is 1/6 BPS. We guess that there is an enhancement of the SUSY when we shrink the loop to the point.

We want to explain why the Wilson loop in the ABJM theory is 1/3 BPS or 1/6 BPS instead of 1/2 BPS.
In the dual IIA string theory side, the Wilson loop
is described by the fundamental string on $AdS_4\times CP^3$ spacetimes. From the supersymmetry analysis \cite{Dru1} of the Killing spinors, it has been known that the fundamental string dual to the straight Wilson loop is not localized at $CP^3$ but smeared along $CP^3$:
the smeared string preserves less of SUSY. 
So, we guess that a similar phenomenon happens for our point-like Wilson loop. 

 \section{BPS operators from the Wilson loop}
In this section, we show that the BPS operators arise in the double series expansion of the point-like Wilson loop operator \eqref{W23} in powers of the loop variables $\delta x^m(s)$ and $\delta M_I{}^{J}(s)=m_{IJ}(s)$ (see also \cite{BD2}).

First, we consider the Wilson loop fluctuated near the point $x^m(s)=x^m_0$.
We parameterize $M_I{}^{J}(s)$ by
\ba M_I{}^J(s)=\begin{pmatrix} 0&0&1&0 \\
0 &0 &0 & 0 \\
0&0&0& 0 \\
0 &0&0&0 \end{pmatrix}+ \begin{pmatrix} m_{11}&m_{12}&0&m_{14} \\
m_{21} &m_{22} &m_{23} & m_{24} \\
m_{31}&m_{32}&m_{33}&m_{34} \\
m_{41} &m_{42}&m_{43}&m_{44} \end{pmatrix}, \label{M218}\ea
where we fix the gauge freedom of the parameter $s$ by imposing $M_1{}^3=1$, and for convenience, omit the label $s$ in $m_{IJ}(s)$. Since the loop coordinates $\delta x^m(s)$ and $m_{IJ}(s)$ should be periodic about $s$, they can be rewritten as follows:
\ba
\delta x^m(s)=\sum ^{\infty}_{n=-\infty}\delta x^{m}_{(n)}e^{2\pi ins/t},\quad 
m_{IJ}(s)=\sum ^{\infty}_{n=-\infty} m_{(n)IJ}e^{2\pi ins/t} \ \text{for} \ (I,J)\neq (3,1). \label{F225}
\ea
In the Fourier space, the loop variables have the zero modes.

The loop variables $\delta x^m(s)$ and $\delta M_I{}^{J}(s)$
are not independent but are related by the BPS condition
 \eqref{B313} and \eqref{B314}.
By substituting \eqref{M218} into the matrix $X_{ab}=(\delta \dot x^2+M^2)_{ab}$ and the BPS condition \eqref{B313} and \eqref{B314}, 
we obtain the following matrix elements and the following equation:
\begin{alignat}{10}
&X_{11}={\delta \dot x}^{2}+{{m_{11}}}^{2}+{m_{12}}{ 
m_{21}}+{m_{31}}+{m_{14}}{m_{41}},\
 X_{12}={m_{11}}{m_{12}}+{m_{12}}{
m_{22}}+{m_{32}}+{m_{14}}{m_{42}}, \notag \\
&X_{13}={m_{11}}+{m_{12}}{m_{23}}+{m_{33}}+{m_{14}}{m_{43}},\
 X_{14}={m_{11}}{m_{14}}+{m_{12}}{ m_{24}}+{m_{34}}+{m_{14}}{m_{44}},
 \notag \\
& X_{21}=m_{21}m_{11}+m_{22}m_{21}+m_{23}m_{31}+m_{24} m_{41},
X_{22}={\delta \dot x}^2+m_{12}{m_{21}}+{{m_{22}}}^{2}+{m_{23}}{ m_{32}}+{m_{24}}{m_{42}}, \notag \\
& X_{23}={m_{21}}+{m_{22}}{m_{23}}+{m_{23}}{m_{33}}+{m_{24}}{m_{43}},\
X_{24}={m_{21}}{m_{14}}+{m_{22}}{m_{24}}+{ 
m_{23}}{m_{34}}+{m_{24}}{m_{44}}, \notag \\  
 &X_{31}={m_{31}}{
m_{11}}+{m_{32}}{m_{21}}+{m_{33}}{m_{31}}+{m_{34}}{m_{41}
},\ X_{32}= {m_{31}}\,{m_{12}}+{m_{32}}{m_{22}}+{m_{33}}{m_{32}}+{m_{34}}{m_{42}}, \notag \\
&X_{33}= {\delta \dot x}^{2}+{m_{31}}+{m_{23}}{m_{32}}+{{m_{33}}}^{2}
+{m_{34}}{m_{43}},\
 X_{34}={m_{31}}{m_{14}}+{m_{32}}{m_{24}}+{ 
m_{33}}{m_{34}}+{m_{34}}{m_{44}}, \notag \\ 
&X_{41}={m_{41}}{
m_{11}}+{m_{42}}{m_{21}}+{m_{43}}{m_{31}}+{m_{44}}{m_{41}
},\  
X_{42}= {m_{41}}{ m_{12}}+{m_{42}}{ m_{22}}+{m_{43}}{m_{32}}+{ m_{44}}{ m_{42}},      \ \notag \\
& X_{43}={m_{41}}+{m_{42}}{m_{23}}+{m_{43}}{m_{33}}+{
m_{44}}{m_{43}} ,
X_{44}= {\delta \dot x}^{2}+{m_{14}}{m_{41}}+{m_{24}}{m_{42}}+
{m_{34}}{ m_{43}}+{{m_{44}}}^{2},      \label{D218} 
\end{alignat}
\ba
&\det X={X_{11}}{X_{22}}{X_{33}}{X_{44}}-{X_{11}}{X_{22}}{ 
X_{34}}{X_{43}}+{X_{11}}{X_{32}}{X_{43}}{X_{24}}-{X_{11}}
{X_{32}}{X_{23}}{X_{44}} \notag \\
&+{X_{11}}{X_{42}}{X_{23}}{X_{34}}-{X_{11}}{X_{42}}{X_{33}}{X_{24}}-{X_{21}}{X_{12}}{X_{33}}{X_{44}}+{X_{21}}{X_{12}}{X_{34}}{X_{43}} \notag \\
&-{X_{21}}{X_{32}}{X_{43}}{X_{14}}+{X_{21}}{X_{32}}{X_{13}}
{X_{44}}-{X_{21}}{X_{42}}{X_{13}}{X_{34}}+{X_{21}}{X_{42}}{X_{33}}{X_{14}}\notag \\
&+{X_{31}}{X_{12}}{X_{23}}{X_{44}}-{X_{31}}{X_{12}}{X_{43}}{X_{24}}+{X_{31}}{X_{22}}
{X_{43}}{X_{14}}-{X_{31}}{X_{22}}{X_{13}}{X_{44}} \notag \\
&+{X_{31}}{X_{42}}{X_{13}}{X_{24}}-{X_{31}}{X_{42}}{X_{23}}
{X_{14}}-{X_{41}}{X_{12}}{X_{23}}{X_{34}}+ 
{X_{41}}{X_{12}}{X_{33}}{X_{24}} \notag \\
&-{X_{41}}{X_{22}}{X_{33}}{
X_{14}}+{X_{41}}{X_{22}}{X_{13}}{X_{34}}-{X_{41}}{X_{32}}
{X_{13}}{X_{24}}+ 
{X_{41}}{X_{32}}{X_{23}}{X_{14}}=0, \label{D422}
\ea
\ba
& m_{31}=\dfrac{4(\delta \dot x)^2+\sum_{(i,j)\neq (3,1)}m_{ij}m_{ji}}{2}.  
\label{M219} 
\ea
The special solution of \eqref{D422} is given by\footnote{We can solve \eqref{B314} about $m_{31}$ since \eqref{B314}
is the equation of second degree about $m_{31}$. However, we cannot expand the denominator of the solution in powers of the loop variables. }
\ba
&  m_{22}=m_{44}=m_{24}=m_{42}=0, \label{M424} \\
&m_{32}=-m_{11}m_{12},\ m_{34}=-m_{11}m_{14},\ m_{41}=-m_{43}m_{33}, \label{M425} \\
&m_{21}=-m_{23}m_{33},\ m_{11}=-m_{33}, \label{M426} \\
&\delta \dot x^2=0, \label{M427} \\
&\Rightarrow  X_{12}=X_{14}=X_{43}=X_{23}=X_{42}=X_{24}=X_{22}=X_{44}=0.  
 \label{M428}
\ea
In the Fourier space, the relations \eqref{M219}$\sim$\eqref{M428} are rewritten as follows:\ba
&m_{32(n)}=-m_{11(n)}m_{12(-n)},\ m_{34(n)}=-m_{11(n)}m_{14(-n)},\
m_{41(n)}=-m_{43(n)}m_{33(-n)}, \label{F427} \\
&m_{21(n)}=-m_{23(n)}m_{33(-n)},\ m_{11(n)}=-m_{33(n)}, \label{F428} \\ 
&\delta x^0_{n}\delta x^0_{-n}=\delta x^1_{n}\delta x^1_{-n}+\delta x^2_{n}\delta x^2_{-n}\ (\text{for $n\neq 0$}), \label{F429} \\ 
&m_{31(n)}=m_{11(n)}m_{11(-n)}. \label{F430}
\ea
Note that while the zero modes $\delta x^m_{(0)}$ are the independent parameters, $\delta x^m_{(n)}$ is not independent.

\vskip5mm

\begin{center}
\begin{tabular}{cccccccc} \hline
 & $m_{13}$ &$m_{12},m_{14},m_{23},m_{43}$ & $\delta x^m,m_{ii},m_{24},m_{42}$ &$m_{21},m_{41},m_{32},m_{34}$ &$m_{31}$ \\ \hline
$J$&1&1/2&0&-1/2&-1 \\ 
$Q(=\Delta -J)$&0&1/2&1&3/2&2 \\ \hline
\end{tabular}
\vskip2mm Table 2: $U_R(1)$ charge of the loop variables $(i=1,2,3,4)$.
\end{center}

\vskip5mm

Recall that the loop variable has the $U_R(1)$ charges as given in Table 2 (see footnote 5). The loop variable also has the dimension $\Delta$ since the dimension of the Wilson loop becomes zero. 
Since the charge $Q(=\Delta -J)$ of the loop variables is well-defined for \eqref{M424}$\sim$\eqref{M428}, we expand the Wilson loop fluctuated at $x^m_0$ up to one in powers of the charge of the loop variable $Q$ and up to one impurity as follows: 
\ba &W[C]=W[C_0]+\int ^t_0ds \sum _{Q(m_{ij})=1/2}m_{ij}(s)\dfrac{\delta W[C]}{\delta m_{ij}(s)}\Biggr|_{C=C_0} \notag \\
&+\int ^t_0ds\delta x^{m}(s)\dfrac{\delta W[C]}{\delta x^m(s)}\Biggr|_{C=C_0}+\int ^t_0ds m_{11}(s)\left(\dfrac{\delta W[C]}{\delta m_{11}(s)}-\dfrac{\delta W[C]}{\delta m_{33}(s)}\right)\Biggr|_{C=C_0} +... \notag \\
   \label{E226}
\ea
where we used \eqref{M424} and \eqref{M426}. Here, $\sum_{Q(m_{kl})=n}$
means that we sum the terms in which the charge of $m_{ij}$ is $n$;
for $n=1/2$, we sum the terms containing $m_{12},m_{14},m_{23}$, and $m_{43}$.
 
We introduce a new functional derivative as follows:
\ba
\dfrac{\delta}{\delta m'}=\dfrac{\delta}{\delta m_{11}}-\dfrac{\delta}{\delta m_{33}}.
\ea 
By substituting the Fourier transformation of the loop variables \eqref{F225} into the resulting formula, we obtain 
\ba
&W[C]=W[C_0]+\sum _{Q(m_{ij})=1/2}\sum_n m_{ij(n)}\int ^t_0ds\dfrac{\delta W[C]}{\delta m_{ij}(s)}\Biggr|_{C=C_0}e^{\f{2\pi ins}{t}} \notag \\
&+\sum_n \delta x^{m}_{(n)}\int ^t_0ds\dfrac{\delta W[C]}{\delta x^m(s)}\Biggr|_{C=C_0}e^{\f{2\pi ins}{t}} 
+ \sum_n m_{11(n)}\int ^t_0ds\dfrac{\delta W[C]}{\delta m'(s)}\Biggr|_{C=C_0} e^{\f{2\pi ins}{t}} +... \notag \\
 \label{E227}
\ea
The functional derivative of \eqref{W23} contained in \eqref{E227} generates the impurity operator (in the sense of the spin chain) as follows \cite{BL1,BL2}:
\ba
&\dfrac{\delta W(C)}{\delta x^{m}(s)} \Biggr| _{C=C_0}=i\mbox{Tr}\left[\left(F_{mn}(x(s))\dot x^{n}(s)+\left(D_{m}Y^IY_{J}^{\dagger}\right) (x(s))M_I{}^J(s)\right)w^{s+t}_s(C)\right]\Biggr| _{C=C_0}  \notag \\
& \qquad\qquad\qquad =i\sum_{J=0}^{\infty}\dfrac{(it)^J}{J!}\mbox{Tr}[\left(D_mA_1B_1\right)(x_0)(A_1B_1(x_0))^J], \label{F25} \\
& \dfrac{\delta W(C)}{\delta M_K{}^L(s)} \Biggr| _{C=C_0}=i\sum_{J=0}^{\infty}\dfrac{(it)^J}{J!}\mbox{Tr}[Y^KY_L^{\dagger}(x_0)(A_1B_1(x_0))^J],  \label{F26}
\ea
where $w_{s'}^{s''}(C)$ describes the Wilson line along the path $C$ from $s=s'$ to $s=s''$ and the covariant derivative $D_m$ including $A_m$ operates on $Y^IY^{\dagger}_J$ so that $Y^IY^{\dagger}_J$ is the adjoint field. Note that \eqref{F25} and \eqref{F26} do not depend on the loop parameter $s$. 
 \begin{center}
\begin{tabular}{cc|c} \hline
  \multicolumn{2}{c|}{ABJM side}& dual IIA superstring side  \\ \hline
 functional derivatives &corresponding operators &  \\ \hline
 $-$&$W(C_0)$& $|0;p^+>$ \\
 $\int dse^{\f{2\pi ins}{t}}\f{\delta}{\delta x^m(s)}$ &$D_mA_1B_1$& $\alpha ^{m+1}_{(0)}$\\
$\int dse^{\f{2\pi ins}{t}}\f{\delta}{\delta m'(s)}$ &$Y^1Y_1^{\dagger}-Y^3Y_3^{\dagger}$& $\alpha ^4_{(0)}$ \\
$\int dse^{\f{2\pi ins}{t}}\f{\delta}{\delta m_{ij}(s)}\Big|_{Q=1/2}$ &$Y^iY_j^{\dagger}$& $\alpha ^k_{(0)}$    \\ \hline
\end{tabular}
\vskip2mm
Table 3: The relation between the functional derivatives and the IIA supergravity modes (the zero modes of the string oscillation operator). In the table, $m=0,1,2$, $(i,j)=(1,2),(1,4),(2,3),(4,3)$ and $k=5,6,7,8$. 
\end{center}

By using \eqref{F25} and \eqref{F26}, we find that the right side of \eqref{E227} becomes the zero mode contribution as follows:
\ba &\sum_{J=0}^{\infty}\dfrac{(it)^J}{J!}\Bigl[\mbox{Tr}[(A_1B_1)^J](x_0)+
it\sum_{Q(m_{ij})=1/2}m_{ij(0)}\mbox{Tr}[Y^iY_j^{\dagger}(A_1B_1)^J](x_0) \notag \\
&+itm_{11(0)}\mbox{Tr}[(Y^1Y_1^{\dagger}-Y^3Y_3^{\dagger})(A_1B_1)^J](x_0)+it\delta x^m_{(0)} \mbox{Tr}[\left(D_mA_1B_1\right)(A_1B_1)^J](x_0)\Bigl].
\label{E228} \ea 
In the higher order of $J$, \eqref{E228} contains the following infinite strings of the operator $A_1B_1$, the bi-fundamental scalars $Y^iY_j^{\dagger}$ and the covariant derivative $D_m$:
\ba
&\mbox{Tr}[(A_1B_1)^J](x_0),\quad \mbox{Tr}[Y^iY_j^{\dagger}(A_1B_1)^J](x_0)
,\quad \mbox{Tr}[\left(D_mA_1B_1\right)(A_1B_1)^J](x_0),  \label{S229} \\
&\mbox{Tr}[(Y^1Y_1^{\dagger}-Y^3Y_3^{\dagger})(A_1B_1)^J](x_0).
\ea
where $(i,j)=(1,2),(1,4),(2,3),(4,3)$ and we separated the operators coupled to the independent loop variables. These BPS operators protected by the supersymmetry correspond to the vacuum state and the excited states of the dual IIA string theory on the pp wave background (see \cite{BB1,BB2}).

By following the correspondence between \eqref{S229} and the dual IIA string excited states, and by following the conjecture in \cite{BD2}, we map the functional derivatives of the Wilson loop to the dual IIA string excited states in Table 3.

Before ending this section, we want to comment on
 the relation between our Taylor expansion of the point-like Wilson loop and the strong coupling expansion: the Wilson loop should be expanded
 in terms of $1/\sqrt{\lambda}$ \cite{Sak1} and have the convergent region of the expansion. 
The operators obtained in the Taylor expansion should be normalized by using their 2-point function.
However, the appearance of the strong coupling expansion is not clear in both our point-like Wilson loop and that of $\mathcal{N}=4$ SYM \cite{BD2}. On the other hand, our Taylor expansion of the Wilson loop \eqref{E228} is similar to the mode expansion of the wave-function in the string field theory (SFT) as was pointed out in the work \cite{BD2}. Since the SFT is independent of the string coupling, \eqref{E228} may be independent of the 't Hooft coupling.
   
\section{Discussions}
In this paper, motivated by the paper \cite{BD2}, we discussed the dual IIA string description of a slightly deformed point-like Wilson loop. 
 By expanding the point-like Wilson loop in powers of the loop variables, we obtained the BPS operators that correspond to the excited string states of the IIA string theory on the pp wave background. 
 Our new result was the impurity operator $Y^1Y_1^{\dagger}-Y^3Y_3^{\dagger}$ that was not determined from the analysis in the gravity side \cite{BB1}. By following the conjecture in \cite{BD2}, we gave the maps from the functional derivatives of the Wilson loop to the dual IIA string excited states in Table 3. 
 
In the same way, the fermionic impurities $\psi_3,\psi^{1\dagger}$ belonging to the BPS multiplets $A_2,\bar{B}_2,\psi_3$ and $\bar{A}_2,B_2,\psi^{1\dagger}$ \cite{BB2} are generated by operating the SUSY transformation $\eta_{IJ}Q^{IJ}$ and $\bar{\eta}^{IJ}\bar{Q}_{IJ}$ on \eqref{W23}, where $I,J=1,..,4$.
Other impurity operators $\psi_2B_1+A_1\psi^{4\dagger}$ and $-\psi_4B_1+A_1\psi^{2\dagger}$ are also generated. The charge $\Delta-J$ of these operators agree with the mass spectrum of the fermions in the gravity side, namely 4 fermions of mass 1/2 and 4 fermions of mass 1.
Thus, we conclude that  8 bosonic impurities and 8 fermionic impurities are obtained by Taylor expansion of the point-like Wilson loop.
 
 The BPS conditions of the Wilson loop \eqref{B313} and \eqref{B314} constrained both the loop variables and the forms of the BPS operators, though we did not prove the uniqueness of the solution of the BPS conditions. In appendix C, we also obtained the BMN operators \cite{BM1} in the higher-order terms of the Taylor expansion \eqref{C60}. It will also be important to compute the anomalous dimension \cite{BI1} of the BMN operators  \eqref{B70} and \eqref{B71}.
 
 Since our Wilson loop is point-like, it is important 
 to construct the straight line BPS Wilson loop that connects the point-like 1/3-BPS Wilson loop to 1/6-BPS Wilson loop in the ABJM theory. 
We leave it in future work.
  
 We also want to comment on the Wilson loop in the ABJM theory from the viewpoint of 11-dimensional theory (M-theory). 
It has been known that in the dual IIA supergravity side, the Wilson loop is described by the
 fundamental string on $AdS_4\times CP^3$ spacetimes. Then, the dual fundamental string should be described by the M2-brane wrapped on the M-circle: the 3-form $C_{(3)}$ in the 11-dimensional supergravity reduces to the NS-NS field $B_{\mu\nu}$ coupling to the IIA string. 
Thus, it is also interesting to analyze the dynamics of the Wilson loop by using the loop equation \cite{Awa1}.\footnote{The loop equation for the pure Chern-Simons theory is slightly different from that of the YM theory.} As the dynamics of the Wilson loop describe the dynamics of the dual string, the loop equation will also describe the dynamics of the M2-brane wrapped on the M-circle. When we analyze the loop equation, however, we need to extend the analysis to include supersymmetry (see also \cite{BL2}).

\vskip2mm
\noindent {\bf Acknowledgments}: 
Especially, we would like to thank M. Fukuma, H. Hata, H. Kawai, and K. Yoshida for the discussions and helpful comments.
We would like to thank K. Katayama for the discussions.
We would like to thank T. Nishioka and T. Takayanagi for the helpful comments on this manuscript.
\vskip2mm

\appendix
\section{Charge conjugation matrix}
In appendix A, we introduce the charge conjugation matrix of both $SO(6)$ and $SO(4)$ and construct the $SO(6)$ reducible representation of the gamma matrices by using the $SO(4)$ reducible representation of the gamma matrices.
 
The charge conjugation matrix constructed by $SO(6)$ gamma matrices satisfies the following relations:
\ba & C\gamma ^iC^{-1}=-\gamma ^{iT},\label{A30} \\
 &C^T=\eta C,  \quad (C\gamma ^{i_1i_2...i_n})^{T}=\gamma ^{i_n T}...\gamma ^{i_1 T}\eta C=\eta (-1)^{\f{1}{2}n(n+1)}C\gamma ^{i_1i_2...i_n}, \label{A31} 
\ea
where \eqref{A30} is the definition of the charge conjugation matrix and $\eta$ is a constant determined later.\footnote{These equations are satisfied for any $SO(N)$.}
\eqref{A31} shows that the matrix defined on the right-hand side of \eqref{A31} transforms as the symmetric tensor or the antisymmetric tensor in terms of the $SO(6)$ spinor index. 
Since the product representation in Table 4 must be consistent with the decomposition $8_s\ \otimes\ 8_s\to 36_S\ \oplus\ 28_A$, $\eta$ must be $+1$. 
\begin{center}
\begin{tabular}{cccccccc} \hline
n & 0 &1 &2 &3 &4 &5 &6 \\ \hline
$_6C_n$&1&6&15&20&15&6&1     \\
$(-1)^{\f{1}{2}n(n+1)}$&$+$&$-$&$-$&$+$&$+$&$-$&$-$ \\ \hline
\end{tabular}
\vskip2mm
Table 4: The product representation of the $8_S$ spinor representation. \\
\end{center}

$C$ defined in \eqref{A30} is the metric of the spinor index as follows:
\ba
&C=C_{\alpha\beta}=C^T,\quad C^{-1}=(C^{-1})^{\alpha\beta},\quad C_{\alpha\beta}(\gamma^i) ^{\beta}{}_{\gamma}=(C\gamma^i)_{\alpha\gamma}, \\
&C\gamma ^i=(C\gamma^i )^T,\quad C_{\alpha\beta}g^{\alpha}{}_{\gamma}g^{\beta}{}_{\delta}=C_{\gamma\delta}, \label{A34}
\ea
where $g$ is the spinor representation of $SO(6)$.\footnote{We can show the second equation in \eqref{A34} by expanding the exponential of $g$ in powers of the generator of the Lorentz algebra.}

We introduce the $SO(4)$ charge conjugation matrix $\hat C$ that satisfies the following relations:
 \ba & \hat C\rho ^k \hat C^{-1}=-(\rho^k)^T, \label{A35} \\
 & \hat C^T=-\hat C. \label{A36} \ea
We decompose the $SO(6)$ Dirac gamma matrices and the charge conjugation matrix as follows:
\ba & \gamma ^k=\begin{pmatrix} 0&\rho^k \\
\rho^k & 0  \end{pmatrix}, \quad  \gamma ^5=\begin{pmatrix} 0&\rho^5 \\
\rho^5 & 0  \end{pmatrix}, \quad  \gamma ^6=\begin{pmatrix} 0&-i \\
i & 0  \end{pmatrix}, \label{M34} \\
&C=\begin{pmatrix} 0&\hat C\ \\
 -\hat C & 0  \end{pmatrix}, \label{M35} \ea
where $\rho^k$ is the $SO(4)$ Dirac gamma matrix. Here, the matrices defined in \eqref{M34} and \eqref{M35} satisfy a Clifford algebra and the definition of the charge conjugation matrix.
\section{Clebsch-Gordan decomposition of the 6 Majorana spinors}
We show the Clebsch-Gordan decomposition of the 6 $SO(1,2)$ Majorana
spinors $\epsilon_i$, which transform as the vector representation of 
$Spin(6)\sim SO(6)$, and show the equality \eqref{C33}. We know 
the following equivalence:
\ba &Spin(6)\ \sim SU(4) \\
     &6_v\ \sim \ 6_A \\
     & 4_s\ \sim \ 4_v  \ea
     
We introduce the new matrix $C_2$ and $4_s$ spinor index $I$ and $J$ as follows:
\ba  &\gamma^i=\begin{pmatrix} 0&\hat\gamma^i _{ I\dot J}  \\
\hat\gamma^i _{ \dot IJ} & 0  \end{pmatrix},\quad
C_2=\begin{pmatrix} 0 & i\hat C\rho_5  \\
-i\hat C\rho_5 & 0  \end{pmatrix}=\begin{pmatrix} 0 & \hat C_{2, I\dot J}  \\
\hat C_{2, \dot I J} & 0  \end{pmatrix} \quad (I,J=1\sim 4), \\
&C_2\gamma ^i=\begin{pmatrix} (\hat C_2\hat \gamma^i) _{ IJ} &0  \\
0 & (\hat C_2\hat \gamma^i)_{\dot I\dot J}  \end{pmatrix},
\ea
By using \eqref{A34}, we can show that $\hat C_2\hat\gamma ^i$ transforms as the antisymmetric tensor in terms of the index $I,J$. 

We describe the Clebsch-Gordan decomposition of $\hat C_2\hat \gamma ^i$ as follows:
\ba &v_{IJ}\to v^i\equiv \sum_{I<J}(\hat C_2\hat \gamma^i)_{IJ}v^{IJ}, \\
 &\omega_{IJ}=\epsilon_i(\hat C_2\hat \gamma^i)_{IJ}. \ea
We choose the basis of the $SO(4)$ gamma matrices as follows:
\ba &\rho^1 =\begin{pmatrix} 0& -i \\
i & 0  \end{pmatrix},\ \rho^2 =\begin{pmatrix} 0& \sigma ^1 \\
\sigma^1 & 0  \end{pmatrix},\ \rho^3 =\begin{pmatrix} 0& \sigma ^3 \\
\sigma ^3 & 0  \end{pmatrix},\ \rho^4 =\begin{pmatrix} 0& \sigma ^2 \\
\sigma ^2 & 0  \end{pmatrix}, 
\ea
\ba
 & \hat C=\rho^2\rho^3,\ \hat C_2=\rho^4\rho^1.
 \ea
After a $SO(6)$ permutation of the $SO(6)$ gamma matrices $\gamma ^i$
including the change of the signs, we obtain the supersymmetry generator $\omega _{IJ}$, which satisfies \eqref{C32} and \eqref{C33} (see also \cite{BJ1}).

It is convenient to introduce the following equations:
\ba &(C\gamma ^i)^*=-\gamma ^iC^*, \\
&(C\gamma ^7)^*=\gamma ^7C^*.  \ea
By using the $SO(6)$ gamma matirces, $C$ and $C_2$,
 we can show \eqref{C33} as follows:
\ba &\epsilon_i( C_2 \gamma^i)
\epsilon^i( C_2^* \gamma^{*i}) \\
 &=\epsilon^i\epsilon_j(C\gamma ^7\gamma ^i\gamma ^7C^*\gamma ^{*j}) \\
&=\epsilon^i\epsilon_j C\gamma ^i\gamma ^jC^*=\epsilon^i\epsilon_i CC^*=\epsilon^i\epsilon_i 1,  \ea
where $\gamma ^7=\gamma ^1\gamma ^2\gamma ^3\gamma ^4$. 

\section{The higher-order terms in the expansion of the point-like Wilson loop}
In this section, we compute the higher-order terms in the expansion \eqref{E226}.
We expand the point-like Wilson loop up to 2 in powers of the charge $Q$ as follows: 
\ba &W[C]=W[C_0]+\int ^t_0ds \sum _{Q(m_{IJ})=1/2}m_{ij}(s)\dfrac{\delta W[C]}{\delta m_{ij}(s)}\Biggr|_{C=C_0} \notag \\
&+\int ^t_0ds\delta x^{m}(s)\dfrac{\delta W[C]}{\delta x^m(s)}\Biggr|_{C=C_0}+\int ^t_0ds \left( m_{11}(s)\dfrac{\delta W[C]}{\delta m_{11}(s)}+m_{33}(s)\dfrac{\delta W[C]}{\delta m_{33}(s)}\right)\Biggr|_{C=C_0}  \notag \\
\ea
\ba
&+\dfrac{1}{2}\int ^t_0ds_1\int ^t_0ds_2\sum _{Q(m_{ij}+m_{kl})=1}m_{ij}(s_1) m_{kl}(s_2)\dfrac{\delta ^2 W[C]}{\delta m_{ij}(s_1)\delta m_{kl}(s_2)}\Biggr|_{C=C_0} \notag  \\
&+\int ^t_0ds \sum_{Q(m_{ij})=3/2}m_{ij}(s)\dfrac{\delta W[C]}{\delta m_{ij}(s)}\Biggr|_{C=C_0} \notag \\
&+\int ^t_0ds_1\int ^t_0ds_2\sum _{Q(m_{ij})=1/2} m_{ij}(s_1)\delta x^n(s_2)\dfrac{\delta ^2 W[C]}{\delta x^n(s_2)\delta m_{ij}(s_1)}\Biggr|_{C=C_0}  \notag \\
&+\dfrac{1}{2}\int ^t_0ds_1\int ^t_0ds_2\sum _{Q(m_{ij}+m_{kl})=3/2}m_{ij}(s_1) m_{kl}(s_2)\dfrac{\delta ^2 W[C]}{\delta m_{ij}(s_1)\delta m_{kl}(s_2)}\Biggr|_{C=C_0} \notag 
\ea
\ba
&+\int ^t_0ds m_{31}(s)\dfrac{\delta W[C]}{\delta m_{31}(s)}\Biggr|_{C=C_0}+
\dfrac{1}{2}\int ^t_0ds_1\int ^t_0ds_2\delta x^{m}(s_1)\delta x^n(s_2)\dfrac{\delta ^2 W[C]}{\delta x^m(s_1)\delta x^n(s_2)}\Biggr|_{C=C_0} \notag \\
&+\int ^t_0ds_1\int ^t_0ds_2\sum _{Q(m_{IJ})=1} m_{IJ}(s_1)\delta x^n(s_2)\dfrac{\delta ^2 W[C]}{\delta x^n(s_2)\delta m_{IJ}(s_1)}\Biggr|_{C=C_0}  \notag \\
&+\dfrac{1}{2}\int ^t_0ds_1\int ^t_0ds_2\sum _{Q(m_{ij}+m_{kl})=2}m_{ij}(s_1) m_{kl}(s_2)\dfrac{\delta ^2 W[C]}{\delta m_{ij}(s_1)\delta m_{kl}(s_2)}\Biggr|_{C=C_0}+
...,
   \label{C60}
\ea
where $\sum_{Q(m_{ij}+m_{kl})=n}$
means that we sum the terms in which the charge of $m_{ij}m_{kl}$ is $n$.
 Recall that the loop variables satisfy the relation \eqref{F427}$\sim$\eqref{F430} as follows:
\ba
&m_{32(n)}=-m_{11(n)}m_{12(-n)},\ m_{34(n)}=-m_{11(n)}m_{14(-n)},\
m_{41(n)}=-m_{43(n)}m_{33(-n)}, \label{C61} \\
&m_{21(n)}=-m_{23(n)}m_{33(-n)},\ m_{11(n)}=-m_{33(n)}, \label{C62} \\ 
&\delta x^0_{n}\delta x^0_{-n}=\delta x^1_{n}\delta x^1_{-n}+\delta x^2_{n}\delta x^2_{-n} \quad (\text{for $n\neq 0$}), \label{C63}  \\
&m_{31(n)}=m_{11(n)}m_{11(-n)}. \label{C65}
\ea 
Because of \eqref{C63}, the spacetime coordinates are twisting for $n\neq 0$.
 By substituting the Fourier transformation of the loop variables \eqref{F225} into \eqref{C60} and by using \eqref{C61}$\sim$\eqref{C65}, we obtain  
\ba
&W[C]=W[C_0]+\sum _{Q(m_{ij})=1/2}\sum_n m_{ij(n)}\int ^t_0ds\dfrac{\delta W[C]}{\delta m_{ij}(s)}\Biggr|_{C=C_0}e^{\f{2\pi ins}{t}} \notag \\
&+\sum_n \delta x^{m}_{(n)}\int ^t_0ds\dfrac{\delta W[C]}{\delta x^m(s)}\Biggr|_{C=C_0}e^{\f{2\pi ins}{t}} 
+ \sum_n m_{11(n)}\int ^t_0ds\dfrac{\delta W[C]}{\delta m'(s)}\Biggr|_{C=C_0} e^{\f{2\pi ins}{t}}  \notag \\
&+\sum _{Q(m_{ij}+m_{kl})=1}\sum _{n_1,n_2}m_{ij(n_1)}m_{kl(n_2)}\int ^t_0ds_1\int ^t_0ds_2\dfrac{\delta ^2 W[C]}{\delta m_{ij}(s_1)\delta m_{kl}(s_2)}\Biggr|_{C=C_0}e^{\f{2\pi i(n_1s_1+n_2s_2)}{t}} \notag 
\ea
\ba
& +\sum _{Q(m_{ij})=1/2}\sum _{n_1,n_2}\delta x^{n}_{(n_1)}m_{ij(n_2)}\int ^t_0ds_1\int ^t_0ds_2\dfrac{\delta ^2 W[C]}{\delta x^n(s_2)\delta m_{ij}(s_1)}\Biggr|_{C=C_0}e^{\f{2\pi i(n_1s_1+n_2s_2)}{t}}   \notag \\
& +\sum _{n_1,n_2}m_{11(n_1)}m_{12(n_2)}\Bigl( \int ^t_0ds_1\int ^t_0ds_2\dfrac{\delta ^2 W[C]}{\delta m'(s_1)\delta m_{12}(s_2)}\Biggr|_{C=C_0}e^{\f{2\pi i(n_1s_1+n_2s_2)}{t}} \notag \\
&-\int ^t_0ds\dfrac{\delta W[C]}{\delta m_{32}(s)}\Biggr|_{C=C_0}e^{\f{2\pi i(n_1+n_2)s}{t}}\Bigr) \notag
\ea
\ba
& +\sum _{n_1,n_2}m_{11(n_1)}m_{14(n_2)}\Bigl( \int ^t_0ds_1\int ^t_0ds_2\dfrac{\delta ^2 W[C]}{\delta m'(s_1)\delta m_{14}(s_2)}\Biggr|_{C=C_0}e^{\f{2\pi i(n_1s_1+n_2s_2)}{t}} \notag \\
&-\int ^t_0ds\dfrac{\delta W[C]}{\delta m_{34}(s)}\Biggr|_{C=C_0}\Bigr)e^{\f{2\pi i(n_1+n_2)s}{t}} \notag \\
& +\sum _{n_1,n_2}m_{11(n_1)}m_{43(n_2)}\Bigl( \int ^t_0ds_1\int ^t_0ds_2\dfrac{\delta ^2 W[C]}{\delta m'(s_1)\delta m_{43}(s_2)}\Biggr|_{C=C_0}e^{\f{2\pi i(n_1s_1+n_2s_2)}{t}} \notag \\
&+\int ^t_0ds\dfrac{\delta W[C]}{\delta m_{41}(s)}\Biggr|_{C=C_0}e^{\f{2\pi i(n_1+n_2)s}{t}}\Bigr) \notag 
\ea
\ba
& +\sum _{n_1,n_2}m_{11(n_1)}m_{23(n_2)}\Bigl( \int ^t_0ds_1\int ^t_0ds_2\dfrac{\delta ^2 W[C]}{\delta m'(s_1)\delta m_{23}(s_2)}\Biggr|_{C=C_0}e^{\f{2\pi i(n_1s_1+n_2s_2)}{t}} \notag \\
&+\int ^t_0ds\dfrac{\delta W[C]}{\delta m_{21}(s)}\Biggr|_{C=C_0}e^{\f{2\pi i(n_1+n_2)s}{t}}\Bigr) \notag \\
& +\dfrac{1}{2}\sum_{m,n}\sum _{n_1,n_2}\delta x^{m}_{(n_1)}\delta x^n_{(n_2)} \int ^t_0ds_1\int ^t_0ds_2 \dfrac{\delta ^2 W[C]}{\delta x^m(s_1)\delta x^n(s_2)}\Biggr|_{C=C_0}e^{\f{2\pi i(n_1s_1+n_2s_2)}{t}}  \notag \\
&+\dfrac{1}{2}\sum _{n_1,n_2}m_{11(n_1)}m_{11(n_2)}\Bigl( \int ^t_0ds_1\int ^t_0ds_2\dfrac{\delta^2W[C] }{\delta m'(s_2)\delta m'(s_1)}\Biggr|_{C=C_0}e^{\f{2\pi i(n_1s_1+n_2s_2)}{t}} \notag \\
&+\int ^t_0ds\dfrac{\delta W[C]}{\delta m_{31}(s)}\Biggr|_{C=C_0}e^{\f{2\pi i(n_1+n_2)s}{t}} \Bigr) +..., 
 \label{C67}
\ea
where the loop variables $\delta x^m_{(n)}$ (for $n\neq 0$) are not independent. Note that in the context of the spin chain, the impurity $D_m$ coupling with $\delta x^m_{(n)}$ mixes with the fermions: we guess that the Wilson loop containing the fermions will be needed to explain the full excited string spectrum in the IIA string theory.
The two functional derivatives of the point-like Wilson loop
are given by 
\ba
&\dfrac{\delta^2 W(C)}{\delta x^{n}(s_2)\delta x^{m}(s_1)} \Biggr| _{C=C_0} \notag \\
&=\dfrac{\delta }{\delta x^{n}(s_2)} i\mbox{Tr}\Bigl[\Bigl( F_{ml}(x(s_1))\dot x^{l}(s_1)+ D_{m}Y^IY_{J}^{\dagger}(x(s_1))M_I{}^J(s_1)\Bigr) w^{s_1+t}_{s_1}(C) \Bigr]\Biggr|_{C=C_0} \notag \\
&=i\mbox{Tr}\left[\delta (s_1-s_2)D_{(n}D_{m)}A_1B_1(x_0)w^{s_1+t}_{s_1}(C_0)\right] +i\mbox{Tr}\left[F_{mn}(x_0)\delta '(s_1-s_2)w^{s_1+t}_{s_1}(C_0) \right] \notag \\
&-\mbox{Tr}\left[D_mA_1B_1(x_0)w_{s_1}^{s_2}(C_0)D_nA_1B_1(x_0)w_{s_2}^{s_1+t}(C_0) \right], \label{C68}
\ea
\ba
&\dfrac{\delta^2 W(C)}{\delta M_K{}^L(s_2)\delta x^{m}(s_1)} \Biggr| _{C=C_0} \notag \\
&=\dfrac{\delta }{\delta M_K{}^L(s_2)} i\mbox{Tr}\Bigl[\Bigl( F_{ml}(x(s_1))\dot x^{l}(s_1)+D_{m}Y^IY_{J}^{\dagger}(x(s_1))M_I{}^J(s_1)\Bigr)w^{s_1+t}_{s_1}(C) \Bigr]\Biggr|_{C=C_0} \notag \\
&=i\mbox{Tr}\left[\delta (s_1-s_2)D_mY^KY_L^{\dagger}(x_0)w^{s_1+t}_{s_1}(C_0)\right] -\mbox{Tr}\Bigl[D_mA_1B_1(x_0)w^{s_2}_{s_1}(C_0)Y^KY^{\dagger}_L(x_0)\cdot \notag \\
&\cdot w^{s_1+l}_{s_2}(C_0) \Bigr],  \label{C69}
\ea
\ba
&\dfrac{\delta^2 W(C)}{\delta M_K{}^L(s_2)\delta M_I{}^J(s_1)} \Biggr| _{C=C_0}=\dfrac{\delta }{\delta M_K{}^L(s_2)} i\mbox{Tr}\left[Y^IY_J^{\dagger}(x(s_1))w^{s_1+t}_{s_1}(C) \right]\Biggr|_{C=C_0} \notag \\
&=-\mbox{Tr}\Bigl[Y^IY_J^{\dagger}(x_0)w^{s_2}_{s_1}(C_0)Y^KY^{\dagger}_L(x_0)w^{s_1+l}_{s_2}(C_0) \Bigr]. \label{C70}
\ea

Next, we introduce the following integral:
\ba &F_2(n,k,J)=\dfrac{1}{(J-k)!k!}\int^1_0 d\tilde{s}\tilde{s}^k (1-\tilde s)^{J-k}e^{2\pi in\tilde s} \notag \\
&\sim \dfrac{1}{JJ!}\exp\left( \f{2\pi ink}{J}\right)\ \text{ (in the large $J$ limit)}. \label{C71} \ea
The derivation of the second line in \eqref{C71} is given in the appendix of \cite{BD2}.
By substituting \eqref{C68}, \eqref{C69} and \eqref{C70} into \eqref{C67} and by transforming the parameter $(s_1,s_2)$ into $(s_1/t,(s_2-s_1)/t)$ for the double integral about $s_1$ and $s_2$, we obtain the local operator expression of the remaining terms as follows:
the $Q=1$ terms in which the product of the loop variables has the charge $Q=1$ become
\ba
&\sum_{Q(m_{ij}+m_{lm})=1}\sum_J\sum_{n_1}(it)^{J+2}m_{ij(n_1)}  m_{lm(-n_1)}\Biggl[\sum_{k=0}^J\mbox{Tr}\Biggl[ Y^iY_j^{\dagger}(A_1B_1)^kY^lY_m^{\dagger}\cdot \notag \\
&\cdot (A_1B_1)^{J-k} \Biggr] (x_0)F_2(n_1,k,J). \label{B442}
\ea
The $Q=3/2$ terms become
\ba
&+\sum_J\sum_{n_1}(it)^{J+2}m_{11(n_1)}m_{12(-n_1)}\Biggl[\sum_{k=0}^J\mbox{Tr}\left[ (Y^1Y_1^{\dagger}-Y^3Y_3^{\dagger})(A_1B_1)^kY^1Y_2^{\dagger}(A_1B_1)^{J-k} \right] (x_0)\cdot \notag \\ &\cdot F_2(n_1,k,J)-\dfrac{1}{(J+1)!}\mbox{Tr}\left[ Y^3Y_2^{\dagger}(A_1B_1)^{J+1}\right] (x_0) \Biggr] \notag \\ 
&+\sum_J\sum_{n_1}(it)^{J+2}m_{11(n_1)}m_{14(-n_1)}\Biggl[\sum_{k=0}^J\mbox{Tr}\left[ (Y^1Y_1^{\dagger}-Y^3Y_3^{\dagger})(A_1B_1)^kY^1Y_4^{\dagger}(A_1B_1)^{J-k} \right] (x_0)\cdot \notag \\ &\cdot F_2(n_1,k,J)-\dfrac{1}{(J+1)!}\mbox{Tr}\left[ Y^3Y_4^{\dagger}(A_1B_1)^{J+1}\right] (x_0) \Biggr] \notag 
\ea
\ba
&+\sum_J\sum_{n_1}(it)^{J+2}m_{11(n_1)}m_{43(-n_1)}\Biggl[\sum_{k=0}^J\mbox{Tr}\left[ (Y^1Y_1^{\dagger}-Y^3Y_3^{\dagger})(A_1B_1)^kY^4Y_3^{\dagger}(A_1B_1)^{J-k} \right] (x_0)\cdot \notag \\ &\cdot F_2(n_1,k,J)+\dfrac{1}{(J+1)!}\mbox{Tr}\left[ Y^4Y_1^{\dagger}(A_1B_1)^{J+1}\right] (x_0) \Biggr] \notag \\ 
&+\sum_J\sum_{n_1}(it)^{J+2}m_{11(n_1)}m_{23(-n_1)}\Biggl[\sum_{k=0}^J\mbox{Tr}\left[ (Y^1Y_1^{\dagger}-Y^3Y_3^{\dagger})(A_1B_1)^kY^2Y_3^{\dagger}(A_1B_1)^{J-k} \right] (x_0)\cdot \notag \\ &\cdot F_2(n_1,k,J)+\dfrac{1}{(J+1)!}\mbox{Tr}\left[ Y^2Y_1^{\dagger}(A_1B_1)^{J+1}\right] (x_0) \Biggr], \label{B443}
\ea
where we have not written the terms dependent on $\delta x^m$ since they are not independent.
We obtain the $Q=2$ terms up to two impurity as follows:
\ba
&+\dfrac{1}{2}\sum_J\sum_{n_1}(it)^{J+2}m_{11(n_1)}m_{11(-n_1)}\Biggl[\sum_{k=0}^J\mbox{Tr}\Bigl[ (Y^1Y_1^{\dagger}-Y^3Y_3^{\dagger})(A_1B_1)^k(Y^1Y_1^{\dagger}-Y^3Y_3^{\dagger})\cdot \notag \\
& \cdot (A_1B_1)^{J-k} \Bigr] (x_0)F_2(n_1,k,J)+\dfrac{1}{(J+1)!}\mbox{Tr}\left[ Y^3Y_1^{\dagger}(A_1B_1)^{J+1}\right] (x_0) \Biggr]+...., \label{B444}
\ea
where we have not written the terms dependent on $\delta x^m$ for the same reason.
Note that $Q=2$ terms also contain the 3 impurity terms.

In the large $J$ limit, \eqref{B442}, \eqref{B443} and \eqref{B444} contain the following BMN operators:
\ba
&\sum_{k=0}^J\mbox{Tr}\left[ Y^iY_j^{\dagger}(A_1B_1)^kY^lY_m^{\dagger}(A_1B_1)^{J-k} \right] e^{\f{2\pi ink}{J}}, \label{B69} 
\ea
\ba
&\sum_{k=0}^J\mbox{Tr}\left[ (Y^1Y_1^{\dagger}-Y^3Y_3^{\dagger})(A_1B_1)^kY^1Y_2^{\dagger}(A_1B_1)^{J-k} \right] e^{\f{2\pi ink}{J}}-\mbox{Tr}\left[ Y^3Y_2^{\dagger}(A_1B_1)^{J+1}\right], \notag \\
&\sum_{k=0}^J\mbox{Tr}\left[ (Y^1Y_1^{\dagger}-Y^3Y_3^{\dagger})(A_1B_1)^kY^1Y_4^{\dagger}(A_1B_1)^{J-k} \right] e^{\f{2\pi ink}{J}}-\mbox{Tr}\left[ Y^3Y_4^{\dagger}(A_1B_1)^{J+1}\right], \notag \\
&\sum_{k=0}^J\mbox{Tr}\left[ (Y^1Y_1^{\dagger}-Y^3Y_3^{\dagger})(A_1B_1)^kY^4Y_3^{\dagger}(A_1B_1)^{J-k} \right] e^{\f{2\pi ink}{J}}+\mbox{Tr}\left[ Y^4Y_1^{\dagger}(A_1B_1)^{J+1}\right], \notag \\
&\sum_{k=0}^J\mbox{Tr}\left[ (Y^1Y_1^{\dagger}-Y^3Y_3^{\dagger})(A_1B_1)^kY^2Y_3^{\dagger}(A_1B_1)^{J-k} \right] e^{\f{2\pi ink}{J}}+\mbox{Tr}\left[ Y^2Y_1^{\dagger}(A_1B_1)^{J+1}\right], \label{B70} 
\ea 
\ba
&\sum_{k=0}^J\mbox{Tr}\left[ (Y^1Y_1^{\dagger}-Y^3Y_3^{\dagger})(A_1B_1)^k(Y^1Y_1^{\dagger}-Y^3Y_3^{\dagger})(A_1B_1)^{J-k} \right] e^{\f{2\pi ink}{J}}+ \notag \\ 
&+\mbox{Tr}\left[ Y^3Y_1^{\dagger}(A_1B_1)^{J+1}\right], \label{B71}
\ea 
 where $(i,j)$ and $(l,m)$ are equal to $(1,2),(1,4),(2,3),(4,3)$.


\begin{thebibliography}{99}
  
\baselineskip=10pt
  
  
\bibitem{BA1}
  J.~M.~Maldacena,
  "The large N limit of superconformal field theories and supergravity,"
  Adv.\ Theor.\ Math.\ Phys.\  {\bf 2} (1998) 231
  [Int.\ J.\ Theor.\ Phys.\  {\bf 38} (1999) 1113]
  [arXiv:hep-th/9711200]; \\
  E.~Witten,
  "Anti-de Sitter space and holography,"
  Adv.\ Theor.\ Math.\ Phys.\  {\bf 2} (1998) 253
  [arXiv:hep-th/9802150]; \\
  S.~S.~Gubser, I.~R.~Klebanov and A.~M.~Polyakov,
  "Gauge theory correlators from non-critical string theory,"
  Phys.\ Lett.\  B {\bf 428} (1998) 105
  [arXiv:hep-th/9802109].
  \bibitem{Dru02}
   N.~Drukker, D.~J.~Gross and H.~Ooguri,
  { "Wilson loops and minimal surfaces,"}
  Phys.\ Rev.\  D {\bf 60}, 125006 (1999)
  [arXiv:hep-th/9904191]. \\
  \bibitem{BS1}
  S.~J.~Rey and J.~T.~Yee,
  { "Macroscopic strings as heavy quarks in large N gauge theory and  anti-de
  Sitter supergravity,"}
  Eur.\ Phys.\ J.\  C {\bf 22}, 379 (2001)
  [arXiv:hep-th/9803001]; \\
  J.~M.~Maldacena,
  { "Wilson loops in large N field theories,"}
  Phys.\ Rev.\ Lett.\  {\bf 80}, 4859 (1998)
  [arXiv:hep-th/9803002]; \\
  N.~Drukker, D.~J.~Gross and A.~A.~Tseytlin,
  { "Green-Schwarz string in AdS(5) x S(5): Semiclassical partition  function,"}
  JHEP {\bf 0004}, 021 (2000)
  [arXiv:hep-th/0001204]; \\
  H.~Kawai and T.~Suyama,
  { "AdS/CFT Correspondence as a Consequence of Scale Invariance,"}
  Nucl.\ Phys.\  B {\bf 789}, 209 (2008)
  [arXiv:0706.1163 [hep-th]]; \\
  T.~Azeyanagi, M.~Hanada, H.~Kawai and Y.~Matsuo,
  "Worldsheet Analysis of Gauge/Gravity Dualities,"
  arXiv:0812.1453 [hep-th].

\bibitem{BP1}
  D.~E.~Berenstein, J.~M.~Maldacena and H.~S.~Nastase,
  "Strings in flat space and pp waves from N = 4 super Yang Mills,"
  JHEP {\bf 0204} (2002) 013
  [arXiv:hep-th/0202021].

\bibitem{BD2}
  A.~Miwa,
  "BMN operators from Wilson loop,"
  JHEP {\bf 0506} (2005) 050
  [arXiv:hep-th/0504039].

 \bibitem{ABJM}
  O.~Aharony, O.~Bergman, D.~L.~Jafferis and J.~Maldacena,
  "N=6 superconformal Chern-Simons-matter theories, M2-branes and their
  gravity duals,"
  JHEP {\bf 0810} (2008) 091
  [arXiv:0806.1218 [hep-th]].
 \bibitem{ABJM2}
  M.~Benna, I.~Klebanov, T.~Klose and M.~Smedback,
  ``Superconformal Chern-Simons Theories and $AdS_4/CFT_3$
 Correspondence,"
  JHEP {\bf 0809} (2008) 072
  [arXiv:0806.1519 [hep-th]].

\bibitem{BB1}
  T.~Nishioka and T.~Takayanagi,
  "On Type IIA Penrose Limit and N=6 Chern-Simons Theories,"
  JHEP {\bf 0808}, 001 (2008)
  [arXiv:0806.3391 [hep-th]].
\bibitem{BB2}  
  D.~Gaiotto, S.~Giombi and X.~Yin,
  "Spin Chains in N=6 Superconformal Chern-Simons-Matter Theory,"
  arXiv:0806.4589 [hep-th].
\bibitem{BB25}
  G.~Grignani, T.~Harmark and M.~Orselli,
  "The SU(2) x SU(2) sector in the string dual of N=6 superconformal
  Chern-Simons theory,"
  Nucl.\ Phys.\  B {\bf 810}, 115 (2009)
  [arXiv:0806.4959 [hep-th]]; \\
\bibitem{BB3}
  O.~Aharony, O.~Bergman and D.~L.~Jafferis,
  "Fractional M2-branes,"
  JHEP {\bf 0811} (2008) 043
  [arXiv:0807.4924 [hep-th]].
\bibitem{BB4}
  M.~Fujita, W.~Li, S.~Ryu and T.~Takayanagi,
  "Fractional Quantum Hall Effect via Holography: Chern-Simons, Edge States,
  and Hierarchy,"
  arXiv:0901.0924 [hep-th]; \\
  D.~Gaiotto and A.~Tomasiello,
  "The gauge dual of Romans mass,"
  arXiv:0901.0969 [hep-th]; \\
  C.~Krishnan, C.~Maccaferri and H.~Singh,
  "Chern-Simons Level Shifts and M2-brane Flows,"
  arXiv:0902.0290 [hep-th].
  
\bibitem{BW1}
  D.~Berenstein and D.~Trancanelli,
  "Three-dimensional N=6 SCFT's and their membrane dynamics,"
  Phys.\ Rev.\  D {\bf 78}, 106009 (2008)
  [arXiv:0808.2503 [hep-th]]; 
  \bibitem{Dru1}
 N.~Drukker, J.~Plefka and D.~Young,
  "Wilson loops in 3-dimensional N=6 supersymmetric Chern-Simons Theory and
  their string theory duals,"
  JHEP {\bf 0811} (2008) 019
  [arXiv:0809.2787 [hep-th]]; 
  \bibitem{BW2}
  B.~Chen and J.~B.~Wu,
  "Supersymmetric Wilson Loops in N=6 Super Chern-Simons-matter theory,"
  arXiv:0809.2863 [hep-th]; \\
  J.~Kluson and K.~L.~Panigrahi,
  "Defects and Wilson Loops in 3d QFT from D-branes in AdS(4) x CP**3,"
  arXiv:0809.3355 [hep-th]; \\
  S.~J.~Rey, T.~Suyama and S.~Yamaguchi,
  "Wilson Loops in Superconformal Chern-Simons Theory and Fundamental Strings
  in Anti-de Sitter Supergravity Dual,"
  arXiv:0809.3786 [hep-th].
 
   \bibitem{BD1}
  A.~M.~Polyakov and V.~S.~Rychkov,
  "Gauge fields - strings duality and the loop equation,"
  Nucl.\ Phys.\  B {\bf 581} (2000) 116
  [arXiv:hep-th/0002106]; \\
  G.~W.~Semenoff and D.~Young,
  "Wavy Wilson line and AdS/CFT,"
  Int.\ J.\ Mod.\ Phys.\  A {\bf 20} (2005) 2833
  [arXiv:hep-th/0405288]; \\
  N.~Drukker and S.~Kawamoto,
  "Small deformations of supersymmetric Wilson loops and open spin-chains,"
  JHEP {\bf 0607} (2006) 024
  [arXiv:hep-th/0604124].
\bibitem{BB0} 
  D.~Astolfi, V.~G.~M.~Puletti, G.~Grignani, T.~Harmark and M.~Orselli,
  "Finite-size corrections in the $SU(2) x SU(2)$ sector of type IIA string
  theory on $AdS_4 x CP^3$,"
  Nucl.\ Phys.\  B {\bf 810}, 150 (2009)
  [arXiv:0807.1527 [hep-th]].
\bibitem{Dba1}
  D.~Bak, D.~Gang and S.~J.~Rey,
  "Integrable Spin Chain of Superconformal U(M)xU(N) Chern-Simons Theory,"
  JHEP {\bf 0810}, 038 (2008)
  [arXiv:0808.0170 [hep-th]].
\bibitem{BJ1}
  S.~Terashima,
  "On M5-branes in N=6 Membrane Action,"
  JHEP {\bf 0808} (2008) 080
  [arXiv:0807.0197 [hep-th]].
\bibitem{BL1}
A.~M.~Polyakov, "Gauge Fields and Strings," Harwood Academic Publishers;
\bibitem{BL2}
  H.~Hata and A.~Miwa,
  "Loop equation in D = 4, N = 4 SYM and string field equation on AdS(5) x
  S**5,"
  Phys.\ Rev.\  D {\bf 73} (2006) 046001
  [arXiv:hep-th/0510150].
\bibitem{Sak1}
  M.~Sakaguchi and K.~Yoshida,
  ``Non-relativistic string and D-branes on AdS(5) x S**5 from semiclassical
  approximation,''
  JHEP {\bf 0705}, 051 (2007)
  [arXiv:hep-th/0703061].
    \bibitem{BM1}
  N.~Beisert, C.~Kristjansen, J.~Plefka, G.~W.~Semenoff and M.~Staudacher,
  "BMN correlators and operator mixing in N = 4 super Yang-Mills theory,"
  Nucl.\ Phys.\  B {\bf 650} (2003) 125
  [arXiv:hep-th/0208178]; \\
  N.~Beisert,
  "BMN operators and superconformal symmetry,"
  Nucl.\ Phys.\  B {\bf 659} (2003) 79
  [arXiv:hep-th/0211032].
  \bibitem{BI1}
    J.~A.~Minahan and K.~Zarembo,
  "The Bethe ansatz for superconformal Chern-Simons,"
  JHEP {\bf 0809} (2008) 040
  [arXiv:0806.3951 [hep-th]]; \\
  D.~Bak, D.~Gang and S.~J.~Rey,
  ``Integrable Spin Chain of Superconformal U(M)xU(N) Chern-Simons Theory,''
  JHEP {\bf 0810}, 038 (2008)
  [arXiv:0808.0170 [hep-th]].
\bibitem{SOC}
  C.~Sochichiu,
  "Dilatation operator in 3d,"
  arXiv:0811.2669 [hep-th].
\bibitem{Awa1}
  M.~A.~Awada,
  "The exact equivalence of Chern-Simons theory with fermionic string theory,"
  Phys.\ Lett.\  B {\bf 221}, 21 (1989).
\end{thebibliography}
\end{document}